\documentstyle[12pt]{article}

\newcommand{\Dbar}{\not{\!{\!D}}}

\begin{document}
\title{The lepton anomaly in the presence of torsion}

\author{Antonio Dobado \\
Departamento de F\'{\i}sica Te\'orica \\
Universidad Complutense de Madrid\\
 28040 Madrid, Spain\\
and\\ Antonio L. Maroto
\\Departamento de F\'{\i}sica Te\'orica \\
Universidad Aut\'onoma de Madrid\\
 28049 Madrid, Spain}

\date{\today}

\maketitle
\begin{abstract}
In the presence of a curved space-time with torsion, the divergence of the
leptonic
current gets new contributions from the curvature and the torsion 
in addition
to those coming from the electroweak fields. However it is 
possible to define a new lepton current which is gauge and 
local Lorentz invariant whose divergence gets no contribution
from the torsion. Therefore we argue that torsion does not
contribute to anomalous lepton number production.\\

\end{abstract}

\newpage
\baselineskip 0.83 true cm
%
%
%

One of the most important open problems in modern physics is that of 
the origin
of the baryon and lepton asymmetries of the universe. From the standard 
big-bang
nucleosynthesis and the present data it is possible to set the  value of
the baryon versus photon densities  in something about $n_B/n_{\gamma}\simeq
10^{-10}$. Since today  the antibaryon density is thought to be completely
negligible $(n_B>>n_{\bar B})$, the above ratio can be considered also 
as the
value of the baryon asymmetry $(\delta)$ before the decoupling of baryons 
and
antibaryons (in fact quarks and antiquarks) and radiation, i.e.
$\delta=(n_B-n_{\bar B})/(n_B+n_{\bar B})\simeq 10^{-10}$. The main 
problem of
any baryogenesis model is to explain the origin of this small number which
however probably gives rise to most of the observed matter around us today.

Sometime ago Sakharov \cite{ZA} set some  necessary ingredients in order 
to get
baryon asymmetry starting from a symmetric universe at early times. This
asymmetry production requires obviously $B$ (baryon number) violation
but also $C$ and $CP$ violation and absence of thermal equilibrium.
The first modern models of baryogenesis appeared last decade with the advent of
Grand Unified Theories (GUT) (see for example \cite{KT} for a review). In 
these
models the decays of the $X$ and $Y$ gauge or the Higgs bosons are $B$
violating. On the other hand $C$ and $CP$ breaking is provided by chiral
interactions and complex Higgs selfcouplings in a natural way. The 
absence of
thermal equilibrium occurs whenever the decay rate of the above mentioned
bosons is smaller than the expansion rate of the universe. 

More recently
it was pointed out that the anomalous electroweak  $L$ (lepton) and 
$B$ numbers
non-conservation present in the Standard Model (SM) \cite{GH}, which 
preserves
$B-L$, could be enough to wash out any $B$ or $L$ asymmetry generated at the 
GUT scale
unless the universe started with some non-vanishing $B-L$ \cite{KRS}. In 
that work the
authors studied also the possibility of baryon asymmetry production by these 
anomalous
electroweak processes themselves but they arrived to the conclusion that 
this is not
possible if the electroweak phase transition is of second order as it 
seems to be the case.

When the gravitational field is taken into account in the computation of
the $L$ and $B$ number anomalies \cite{Sal}, one observes that $B-L$ is 
not conserved 
any more. Thus
one could speculate on the possibility that these gravitational 
effects could 
generate a
$B-L$ asymmetry and that the usual electroweak effect would just 
redistribute
 the relative
amounts of $B$ and $L$. However, it is not obvious how to make such an 
scenario to work in
detail \cite{I}.  

In this work we will consider another potential source of anomalous $B-L$ 
non-conservation which
could appear when, in addition to the gravitational field, we have also 
torsion. As it is well
known, once a metric is given, there  is a unique affine connection which 
is metric
compatible and torsion free. This is the Levi-Civita connection which is 
given by the
Christoffel symbols and it was the only one considered by Einstein in his 
original
formulation of General Relativity. However, in all the modern discussions 
on gravitation,
the metric (the vierbein) and the connection are treated as independent 
structures and then
torsion appears in a natural way. In particular,  whenever fermions are 
present, as it happens in the SM, the torsion is different from 
zero \cite{Hehl}.  
In addition, when one
considers only minimal couplings, fermions are the only SM particles which 
couple to the
torsion, in particular, to its pseudo-trace. In a recent work \cite{DoMa96} 
the authors
have computed the $B$ and $L$ anomaly for the SM defined on a gravitational
field  with torsion by which we mean a metric (vierbein) and an arbitrary
(but metric compatible) affine connection. We have found that, in addition 
to the electroweak
and curvature terms, there are also new contributions coming from torsion 
to the $L$
anomaly. Thus, any amount of torsion present in the background gravitational 
field can in
principle be transformed in lepton number. This 
mechanism for the $B-L$ production could be used to build a new scenario of  
lepton asymmetry generation. The resulting lepton asymmetry could be 
transformed 
again by the more usual electroweak anomaly  effects in a baryon 
asymmetry. In the 
following we will duscuss that issue in detail and we will conclude that, 
due to the possibility of redefining a gauge and local Lorentz 
invariant lepton current in presence of torsion, no real 
anomalous lepton production takes place from the above mechanism.

The Euclidean SM matter lagrangian in a curved space-time with torsion can 
be written
for one family as follows \cite{DoMa96}:
\begin{eqnarray}
{\cal L}_m=Q^{\dagger}{{\Dbar}^Q}Q+L^{\dagger}{{\Dbar}^L}L
\label{lm}
\end{eqnarray}
where the matter fields appear in doublets:
\begin{eqnarray}
Q=\left[
\begin{array}{c}
u\\
d
\end{array}
\right]
\;L=\left[
\begin{array}{c}
\nu\\
e
\end{array}
\right]
\end{eqnarray}
The Dirac operators for quarks and leptons are given by
\begin{eqnarray}
i{\Dbar}^Q & = & i\gamma^{\mu}(\partial_{\mu}+{\bf
\Omega}^{Q}_{\mu}+{\bf G}_{\mu}+{\bf W}^{Q}_
{\mu}P_L+ {\bf B}^{Q}_{\mu}+{\bf S}^{Q}_{\mu}\gamma_5) \nonumber \\
i{\Dbar}^L & = & i\gamma^{\mu}(\partial_{\mu}+{\bf \Omega}^{L}_{\mu}
  +{\bf W}^{L}_{\mu}P_L+
{\bf B}^{L}_{\mu}+{\bf S}^{L}_{\mu}\gamma_5)
 \label{dos}
 \end{eqnarray}
Here ${\bf G}_{\mu}=-ig_SG_{\mu}^a\Lambda^a$, 
${\bf W}_{\mu}=-igW_\mu^aT^a$ and 
${\bf B}_{\mu}=ig'B_\mu({\bf y}_LP_L+{\bf y}_R P_R)$ are the 
usual gauge fields 
corresponding to the $SU(3)_c$, $SU(2)_L$ and $U(1)_Y$ groups respectively 
and the
hypercharge matrices are given by
\begin{eqnarray}
{\bf y}^{Q,L}_{L,R}=
\left(
\begin{array}{cc}
y_{L,R}^{u,\nu} & \; \\
\; & y_{L,R}^{d,e}
\end{array}
\right) 
\label{hym} 
\end{eqnarray}
 In addition we
have the Levi-Civita spin connection and torsion terms that 
for leptons read:
\begin{eqnarray}
{\bf \Omega}^L_{\mu}=-\frac{i}{2}\Gamma^{a\;b}_{\;\mu}\left(
\begin{array}{cc}
P_L\Sigma_{ab} & \; \\
\; & \Sigma_{ab}
\end{array} 
\right) \; ,
{\bf S}^L_{\mu}\gamma_5=-\frac{1}{8}S_{\mu}\left(
\begin{array}{cc}
P_L\gamma_5 & \; \\
\; &\gamma_5
\end{array} 
\right)
\label{spin} 
\end{eqnarray}
where $\Sigma_{ab}$ are the $SO(4)$ generators in the spinor 
representation, $\{\Gamma^{a\;b}_{\;\mu} \}$ are the components of the 
Levi-Civita spin connection and 
$S_\alpha=\epsilon_{\mu\nu\lambda\alpha}T^{\mu\nu\lambda}$ is the 
pseudo-trace of the torsion tensor. Such a tensor is defined as follows:
given a metric connection $\{\hat\Gamma^{a\;b}_{\;\mu} \}$, 
it can be written in general as:
\begin{eqnarray}
\hat\Gamma^{a\;b}_{\;\mu}=\Gamma^{a\;b}_{\;\mu}+
e^a_{\nu}e^{\lambda b}K^{\nu}_{\;\mu\lambda}
\label{tors}
\end{eqnarray} 
where $e^a_{\nu}$ is the vierbein and $K^{\nu}_{\;\mu\lambda}$ is known 
as the 
contorsion tensor
which 
 is related to the torsion tensor by \cite{Nakahara}:
\begin{eqnarray}
 K^{\nu\mu\lambda}=\frac{1}{2}(T^{\nu\mu\lambda}+T^{\mu\nu\lambda}+
T^{\lambda\nu\mu})
\end{eqnarray}

Since there are no right neutrinos, the upper components in Eq.\ref{spin} 
appear 
multiplied by a left projector. 
In the case of quarks such projectors are not present. As is well known, the 
above lagrangian
is invariant under the following global transformations:
\begin{eqnarray}
Q\rightarrow Q+i\alpha Q \; , L\rightarrow L + i\beta L
\end{eqnarray}
The corresponding Noether currents are nothing but the baryon and lepton 
number 
currents:
\begin{eqnarray}
j^\mu_B=\frac{1}{N_c}Q^\dagger \gamma^\mu Q \; 
, j^\mu_L=L^\dagger \gamma^\mu L
\end{eqnarray}
which classically are conserved, i.e $\nabla_\mu j^\mu_{Q,L}=0$, where 
$\nabla_{\mu}$ is
the Levi-Civita covariant derivative. However these conservation laws are 
violated due to
quantum effects. The corresponding anomalies can be obtained using, for 
instance, the
Fujikawa method based on the $i{\Dbar}^\dagger i{\Dbar}$ 
and $i{\Dbar} i{\Dbar}^\dagger$ operators
\cite{Fu}. The gaussian regulators associated to these operators respect 
the gauge and 
local Lorentz symmetries of the theory. This procedure yields for 
the anomalies the 
following results (for Euclidean signature) \cite{DoMa96}:
\begin{eqnarray}
\nabla_{\mu}j^{\mu}_B=\frac{1}{32\pi^2}\epsilon^{\mu \nu \alpha
\beta}\left(\frac{g^2}{2} W^a_{\mu \nu}
W^a_{\alpha\beta}+g'^2B_{\mu\nu}B_{\alpha\beta}\sum_{u,d}
(y_L^2-y_R^2)\right)
\label{ban}
\end{eqnarray}
and 
\begin{eqnarray}
\nabla_{\mu}j^{\mu}_L=\frac{1}{32\pi^2}\left \{ -\frac{\epsilon^{\alpha 
\beta
\gamma \delta}}{24} R_{\mu \nu \alpha \beta}
R^{\mu\nu}_{\; \; \;\gamma\delta}+\frac{\epsilon^{\alpha \beta
\gamma \delta}}{48}S_{\beta ; \gamma}S_{\delta ;
\alpha} +\epsilon^{\alpha \beta
\gamma \delta}\left(\frac{g^2}{2}W^a_{\gamma \delta}
W^a_{\alpha\beta}\right. \right. \nonumber \\
\left. \left.+g'^2B_{\gamma
\delta}B_{\alpha\beta}\sum_{\nu,e}(y_L^2-y_R^2)\right)+
\frac{1}{6} \Box S^{\alpha}_{\; ;\alpha}+ \frac{1}{96}\left(S^{\alpha}
S^{\nu}S_{\alpha}\right)_{;\nu} -\frac{1}{6}\left(R^{\nu
\alpha}S_{\alpha}-\frac{1}{2}RS^{\nu}\right)_{; \nu}\right\}
\label{lan}
\end{eqnarray}
These expressions are gauge and local Lorentz invariant. 
In principle, it is possible to use a different regulator within 
the Fujikawa 
method and,
in fact,  the $(i{\Dbar})^2$ 
operator is also frequently used in the calculation of the so called 
consistent gauge
anomaly. 
This regulator yields no anomaly in $L$ and $B$ but its 
eigenvalues are not gauge invariant. The choice of regulator have the same 
effects
as the choice of the shifted momenta in the divergent integrals appearing in 
the triangle
diagrams  contributing to the axial anomaly. In that case, it is possible to 
choose the
integrated momenta in  such a way that the axial anomaly vanishes, but this 
introduces an
anomaly in the gauge current. Conversely, it is possible to cancel the gauge 
anomaly but
maintaining the  axial  one. However axial anomalies have 
measurable physical 
effects as
the
$\pi^0$ decay into  two photons and therefore the above
ambiguities in  their precise 
value should be removed. This can be achieved by demanding that gauge 
invariance is 
respected in the calculations which in turn 
imposes the use of a gauge and local Lorentz invariant regulator as the one 
considered here to
obtain Eq.\ref{ban} and Eq.\ref{lan}.

The resulting  lepton anomaly has terms depending on the curvature and the 
torsion that appear due to the absence of one of the chirallity components 
of the neutrino
field. Such terms are not present in the baryonic anomaly since quarks have 
both chirallity
components. Thus in contrast with flat space-time, $B-L$  is spoiled in 
the presence 
of curvature and in principle also in presence of torsion. 

Therefore we have three different kinds of potential contributions 
to the lepton 
anomaly. First we
have a $\epsilon^{\alpha \beta
\gamma \delta} R_{\mu \nu \alpha \beta}
R^{\mu\nu}_{\; \; \;\gamma\delta}$ term where 
$R^{\mu\nu}_{\; \; \;\gamma\delta}$ is the 
curvature associated to the Levi-Civita spin connection. This term 
integrated over the
whole space-time is related to a topological invariant called the Pontryagin 
index of the manifold
\cite{Eg} and it can be written as a four-divergence, i.e, 
$\nabla_{\mu}K^\mu=\epsilon^{\alpha \beta
\gamma \delta}/(768\pi^2) R_{\mu \nu \alpha \beta}
R^{\mu\nu}_{\; \; \;\gamma\delta}$. This fact allows us to redefine the 
lepton current as
$\tilde j^\mu_L=j^\mu_L+K^\mu$. In absence of the other two pieces of the 
anomaly, this 
current would be conserved, although it is not local Lorentz invariant. Such 
a topological 
term could give rise to actual contributions to the $L$ violation through 
the so called 
gravitational instantons \cite{EgFr} that would be relevant in the 
context of quantum 
gravity and, as mentioned above, its possible relevance for the 
generation of
asymmetries was sketched in \cite{I}. Second, there is also the 
well known $SU(2)_L$ and
$U(1)_Y$ gauge contributions to the anomaly which were studied 
in a cosmological context in
\cite{KRS}. Finally, we also have the possible torsion contributions in 
which we are interested in
this work. 

Concerning to the possibility of having anomalous lepton 
production from torsion two important facts should be 
mentioned. First one could think,  following the usual 
topological interpretation of the
axial anomalies, that the lepton number anomaly could be given
 by the index of the Dirac
operator with a spin connection with torsion. In other words: 
$\nabla_\mu j^\mu_L= (\epsilon^{\alpha
\beta \gamma \delta}/(768\pi^2) \hat R_{\mu \nu \alpha \beta}
\hat R^{\mu\nu}_{\; \; \;\gamma\delta}+\{\mbox{electroweak gauge terms}\})$ 
where 
$\hat R_{\mu \nu \alpha \beta}$ is the curvature associated to the 
full connection (with 
torsion). However it can be explicitly shown that this is not the case since
for a completely antisymmetric torsion we have
\begin{eqnarray}
\frac{\epsilon^{\alpha \beta
\gamma \delta}}{24} \hat R_{\mu \nu \alpha \beta}
\hat R^{\mu\nu}_{\; \; \;\gamma\delta}&=&\frac{\epsilon^{\alpha \beta
\gamma \delta}}{24} R_{\mu \nu \alpha \beta} 
R^{\mu\nu}_{\; \; \;\gamma\delta}
+\frac{1}{432}\epsilon^{\alpha\beta\gamma\delta}S_{\alpha ; \beta}
S_{\gamma ; \delta}\nonumber \\
&+&\frac{1}{6}\left(\frac{1}{3}R^{\beta\alpha}S_{\alpha ; \beta}
-\frac{1}{6}S^\mu_{\;\; ;\mu}R
-\frac{1}{432}S^2 S^\mu_{\;\; ;\mu}-\frac{1}{216}S_{\alpha ; \beta}
S^{\alpha}S^{\beta}\right)
\end{eqnarray}
and this expression does not agree with the result in Eq.\ref{lan}.

Second, and more relevant is the fact that the  
torsion contribution  to the anomaly is a four-divergence and it can 
be absorbed  in the
redefinition of the  lepton current as follows:
\begin{eqnarray}
\tilde j^{\mu}_L=j^{\mu}_L
-\frac{1}{32\pi^2} \left( \frac{1}{6}S_{\alpha}^{\; ;\alpha\mu}
+\frac{1}{96}S^{\alpha}S^{\mu}S_{\alpha}
-\frac{1}{6}\left(R^{\mu\alpha}S_{\alpha}
-\frac{1}{2}RS^{\mu}\right)
+\frac{1}{48}\epsilon^{\mu\beta\gamma\delta}S_{\beta;\gamma}S_{\delta}
\right)  
\label{redcur}
\end{eqnarray} 
The new current $\tilde j^{\mu}_L$ is still gauge and local 
Lorentz invariant 
and depends explicitly on the
torsion. In the absence of the other contributions to the anomaly this 
current is 
conserved, that is, 
$\nabla_{\mu}\tilde j^{\mu}_L=0$. Thus we observe that the possible 
effects of the torsion can be 
eliminated away by this redefiniton of the 
lepton current. Notice that the new definiton, being gauge and 
local Lorentz invariant, is physically meaningful and should 
satisfy the appropriate Ward identities for a properly 
defined lepton current.

Let us stress once again the main differences between the 
torsion terms and the gauge and
gravitational contributions to the lepton anomaly. First, as we have just 
mentioned, the
redefined lepton current in Eq.\ref{redcur} is gauge and local 
Lorentz invariant, moreover there is no local
symmetry associated to torsion that is spoiled
in the redefinition. However when one absorbes in a similar way 
the gauge and gravitational terms, the new lepton current  
no longer respects gauge and Lorentz symmetries. In addition, the 
gauge and gravitational
terms appear in the form of topological invariants. This fact implies that
the variation of the lepton number is related to quantum 
tunneling processes between different classical vacua 
characterized by different winding numbers. Such processes, as it is
well known, are dominated by instantons in the semi-classical approximation. 
However, as we mentioned before, the torsion contribution is 
not topological. In fact it has been shown \cite{Nieh} that apart from 
the Pontryagin and Euler classes, there is only one more non-trivial 4-form 
containing torsion and such form does not agree with 
the result in Eq.\ref{lan}.

An alternative way to show that torsion does not contribute to the
lepton anomaly is based on a different choice of regulator. In fact
let us take as regulators for the anomalies the operators 
$i\bar{\Dbar}^\dagger i\bar{\Dbar}$ 
and $i\bar{\Dbar} i\bar{\Dbar}^\dagger$, where $\bar{\Dbar}$ is
the ${\Dbar}$ operator in which the torsion field has been set to
zero. These operators respect all the gauge and local Lorentz
symmetries of the theory and therefore they are also valid
as regulators. However they do not depend on torsion, this implies 
that the regulated anomalies cannot depend on torsion, as expected.

To summarize, it is shown that, although there are torsion contributions
to the lepton anomaly (which has been obtained by means of a 
gauge and local Lorentz invariant regulator), however such terms cannot
give rise to lepton number generation.

\vskip 1.0cm

{\bf Acknowledgments:} 
This work has been supported in part by the Ministerio de Educaci\'on y
Ciencia (Spain) (CICYT AEN96-1634). 


\end{document}